\documentclass[apj,iop]{emulateapj}  

\usepackage{graphicx} 
\usepackage{apjfonts} 

\shorttitle{Contribution of the solar internetwork to the network
  magnetic flux} 

\shortauthors{Go\v{s}i\'{c} et al.}

\begin{document}

\title{The Solar Internetwork. I. Contribution to the Network Magnetic Flux}

\author{M.~Go\v{s}i\'{c}$^1$}
\author{L.~R.~Bellot Rubio$^1$}
\author{D.~Orozco Su\'arez$^2$}
\author{Y.~Katsukawa$^3$}
\author{J.~C.~del Toro Iniesta$^1$}

\affil{$^1$ Instituto de Astrof\'{\i}sica de Andaluc\'{\i}a (CSIC),
  Apdo.\ 3004, 18080 Granada, Spain; mgosic@iaa.es}
\affil{$^2$Instituto de Astrof\'isica de Canarias, E-38205 La Laguna, Tenerife, Spain}
\affil{$^3$ National Astronomical Observatory of Japan, 2-21-1 Osawa, Mitaka, Tokyo 181-8588, Japan}

\begin{abstract}

The magnetic network observed on the solar surface harbors a sizable fraction 
of the total quiet Sun flux. However, its origin and maintenance are not well 
known. Here we investigate the contribution of internetwork magnetic fields to 
the network flux. Internetwork fields permeate the interior of supergranular 
cells and show large emergence rates. We use long-duration sequences of 
magnetograms acquired by Hinode and an automatic feature tracking algorithm to
follow the evolution of network and internetwork flux elements. We find that 
14\% of the quiet Sun flux is in the form of internetwork fields, with little 
temporal variations. Internetwork elements interact with network patches and 
modify the flux budget of the network, either by adding flux (through merging 
processes) or by removing it (through cancellation events). Mergings appear to 
be dominant, so the net flux contribution of the internetwork is positive. The 
observed rate of flux transfer to the network is 
$1.5\times10^{24}$~Mx~day$^{-1}$ over the entire solar surface. Thus, 
the internetwork supplies as much flux as is present in the network in only 
9-13 hours. Taking into account that not all the transferred flux is 
incorporated into the network, we find that the internetwork would be able to 
replace the entire network flux in approximately 18--24 hours. This renders the 
internetwork the most important contributor to the network, challenging the 
view that ephemeral regions are the main source of flux in the quiet Sun. About 
40\% of the total internetwork flux eventually ends up in the network.

\end{abstract}

\keywords{Sun: magnetic field -- Sun: photosphere}

\section{Introduction}

The quiet Sun (QS) consists of network (NE) and internetwork (IN) regions. The 
photospheric NE outline the borders of supergranular cells, while the IN 
represents the cell interiors. Both are permeated by magnetic fields. 
These fields are strong (of order 1 kG) in the NE and much weaker (of order 1 
hG) in the IN (for reviews see \citealt{Solanki}, \citealt{DeWijn}, 
\citealt{2011ASPC..437..451S}, and \citealt{BellotOrozco}). Because of their 
abundance, QS magnetic fields play a crucial role in the energy budget of the 
solar atmosphere. Thus, it is important to understand their nature and how they 
are maintained on the solar surface.

An important open question is the origin of the NE. In polarized light, NE 
patches are the most prominent structures of the QS. They live for hours to 
days and carry a total flux of $10^{23}-10^{24}$~Mx over the entire solar 
surface \citep{Simon2001,Hagenaar2003, Hagenaar2008, Zhou2013}. This is 
comparable to the flux of active regions \citep[$\sim8 \times 10^{23}$~Mx at 
solar maximum;][]{1994SoPh..150....1S}. Despite its importance, we know little 
about how the NE is formed and sustained. Since the NE encloses IN regions, it 
is natural to expect a significant contribution of the IN to the NE flux. 

IN fields were discovered by \citet{LivingstonHarvey}. In high-resolution 
magnetograms, they are observed as small, isolated features with fluxes in the 
range $10^{16}$ to $10^{18}$~Mx that continually appear in the interior of 
supergranular cells and move toward the NE \citep[e.g.,][]{Zirin1985, Wang, 
Centeno, Lamb2008, Orozco2008, MartinezLuis, Lamb2010, Zhou2013}. High 
sensitivity is needed to detect them because they produce tiny polarization 
signals. For this reason their properties are not well known. However, there 
are indications that, at any time, the weak IN elements may carry up to half of 
the QS magnetic flux \citep{Wang,Meunier,Lites2002,Zhou2013}, which testifies 
to their importance. The flux appearance rates inferred in the IN vary from the 
$10^{24}$~Mx~day$^{-1}$ of \cite{Zirin1987} to the $3 
\times 10^{25}$~Mx~day$^{-1}$ of \cite{ThorntonParnell} and the $3 \times
10^{26}$~Mx~day$^{-1}$ of \cite{Zhou2013}. These values imply that small-scale 
IN features bring to the surface considerably more flux than active regions 
during the maximum of the solar cycle \citep[$6 \times 
10^{21}$~Mx~day$^{-1}$;][]{1994SoPh..150....1S}. IN elements have mean 
lifetimes of less than 10 minutes \citep{Zhou2010,milan} and many never leave 
the IN, but others survive long enough to reach the NE and deposit their flux 
there \citep{LivingstonHarvey, Zirin1985, WangZirin, Orozco2012}. How exactly 
they contribute to the NE flux is still unknown.

The current paradigm is that ephemeral regions (ERs), and not IN fields, are 
the primary source of flux for the NE. Discovered by \cite{HarveyMartin}, ERs 
are bipolar features with fluxes in the range $5-30 \times 10^{18}$~Mx and 
lifetimes of 3--4 hours \citep{Harvey1975, Title, Chae, Hagenaar2001, 
Hagenaar2003, Hagenaar2008}. They can be considered as the largest and 
strongest IN features, comparable to individual NE patches. \cite{Martin1990}
stated without proof that 90\% of the NE flux concentrations come from ERs, 
while the weaker IN elements contribute only 10\% of the NE flux.

\cite{Schrijver} developed a set of equations to model the distribution and 
evolution of the NE flux taking into account the balance between flux 
emergence, fragmentation, merging, and cancellation. Based on \cite{Martin1990} 
claim, they neglected IN elements as a source of flux for the NE and only 
considered ERs, determining their flux emergence rate to be $7.2 \times
10^{22}$~Mx~day$^{-1}$ from data taken by the Michelson Doppler Imager
(MDI) aboard the SOHO satellite \citep{Scherrer}. \cite{Schrijver}
concluded that bipolar ERs alone can explain the flux of the NE, 
with only a small fraction of the NE originating from active regions.
Indeed they found that the total NE flux could be replaced by newly
emerged ERs in 1.5--3 days. The ratio of the total NE flux to the ER
flux appearance rate also gave a flux replacement time of 72 hours.
Using a kinematic model, \cite{Simon2001} studied the transport of ER
fragments by granular and supergranular flows and verified that the
injection of bipolar ERs at the observed rates would lead to a
statistically stable NE. \cite{Hagenaar2003} confirmed the importance
of ERs for the maintenance of the NE and inferred flux replacement
times of 8--19 hours, also using MDI observations. Later,
\cite{Hagenaar2008} revised this value down to only 1--2 hours based
on high-resolution MDI magnetograms.

Despite the large differences in the reported flux replacement times,
all authors seem to agree that ERs are the main source of flux for the
NE. However, most of the analyses neglect the contribution of the weak
IN elements---those with fluxes below $10^{18}$~Mx---, because the
sensitivity and spatial resolution of MDI is insufficient to detect
them. Thus, a potentially important source of flux is not accounted
for. \cite{Lamb2008} suggested that the IN elements hidden to MDI may
actually supply more flux to the NE than ERs.

In this paper we determine the contribution of the IN to the NE using
very long time sequences of magnetograms recorded with the Narrowband
Filter Imager \citep[NFI;][]{Tsuneta} aboard the Hinode satellite
\citep{2007SoPh..243....3K}. Our observations are ideally suited to
investigate the transfer of flux from the IN to the NE because they
show the evolution of IN elements and their interaction with the NE on
time scales from minutes to days, covering most of the lifetime of
supergranular cells. Thus, we can study the flux history of individual
cells and the network regions surrounding them at unprecedented
sensitivity, spatial resolution, and cadence. All these ingredients
are needed to determine the instantaneous flux that is transferred
from the IN to the NE and how it varies with time. In Paper II of
this series we will study in detail the rates of flux appearance
and disapperance in the IN.

After describing the observations in Sect.~\ref{sect2}, we present our
strategy to track QS magnetic flux patches and follow their
interactions, explaining the different processes through which IN
elements contribute flux to the NE (Sect.~\ref{method}). In
Sect.~\ref{results} we determine the fraction of QS flux that is in
the form of IN elements, the temporal evolution of the total IN and NE
fluxes, the rates at which IN flux is transferred to the NE, the
fraction of the IN contribution to the NE due to ERs, and the fraction
of the total IN flux that eventually ends up in the NE.

\section{Observations and data processing}
\label{sect2}

The observations analyzed here were recorded on 2010 January 20--21
(data set 1) and 2010 November 2--3 (data set 2) as part of the
Hinode Operation Plan 151 entitled {\em ``Flux replacement in the photospheric 
network and internetwork''}. The Hinode NFI took
shutterless Stokes I and V filtergrams of the QS at disk center in the
two wings of the \ion{Na}{1} D1 589.6~nm line, $\pm$16~pm away from
its core. Eight $I\pm V$ image pairs were added to get I and V at each
wavelength position. With individual exposure times of 0.2~s, this
implies a total integration time of 6.4~s per line scan.
Three contiguous regions were recorded consecutively and processed
together to make up a total field of view (FOV) of
$93\arcsec\times123\arcsec$ (data set 1) and
$93\arcsec\times82\arcsec$ (data set 2). In both cases the pixel size
was 0\farcs16 and the spatial resolution about 0\farcs3. The duration
of data set 1 is 20~h, with two interruptions of 6 and 29 min caused
by transmission problems and/or data recorder overflows. The second data set 
lasts for 38~h and has two gaps of 1~h and 31 min. The cadence of the 
observations is 60 and 90~s, respectively. These and other details of the 
measurements are given in Table~\ref{data}.

\begin{deluxetable}{lcccc}
  \tablecolumns{3} \tablewidth{0pc} \tablecaption{HOP 151 data sets
    details} \tablehead{ \colhead{} & \colhead{Set 1} & \colhead{Set
      2}} \startdata
  Date & 2010 Jan 20--21 & 2010 Nov 02--03 \\
  Starting time & 13:56:34 UT& 04:02:34 UT\\
  Duration (hh:mm:ss) & 20:07:00 & 38:25:30 \\
  Cadence (s) & 60 & 90 \\
  Exposure time per magnetogram (s) & 6.4 & 6.4 \\
  Noise (Mx cm$^{-2}$) & 4 & 4 \\
  Pixel size  & 0\farcs16  & 0\farcs16 \\
  Total FOV & $93\arcsec\times123\arcsec$ & $93\arcsec\times82\arcsec$ \\
  Common FOV  & $82\arcsec\times113\arcsec$ & $80\arcsec\times74$\arcsec \\
  Number of gaps & 2 & 2 \\
  Duration of gaps & 00:06:00 & 01:01:30 \\
  & 00:29:11 & 00:31:31 \\
  Longest interval without gaps & 11:11:00 & 24:42:00
\enddata
\label{data}
\end{deluxetable}

\begin{figure*}[t]
\begin{center}
\resizebox{1\hsize}{!}{\includegraphics[]{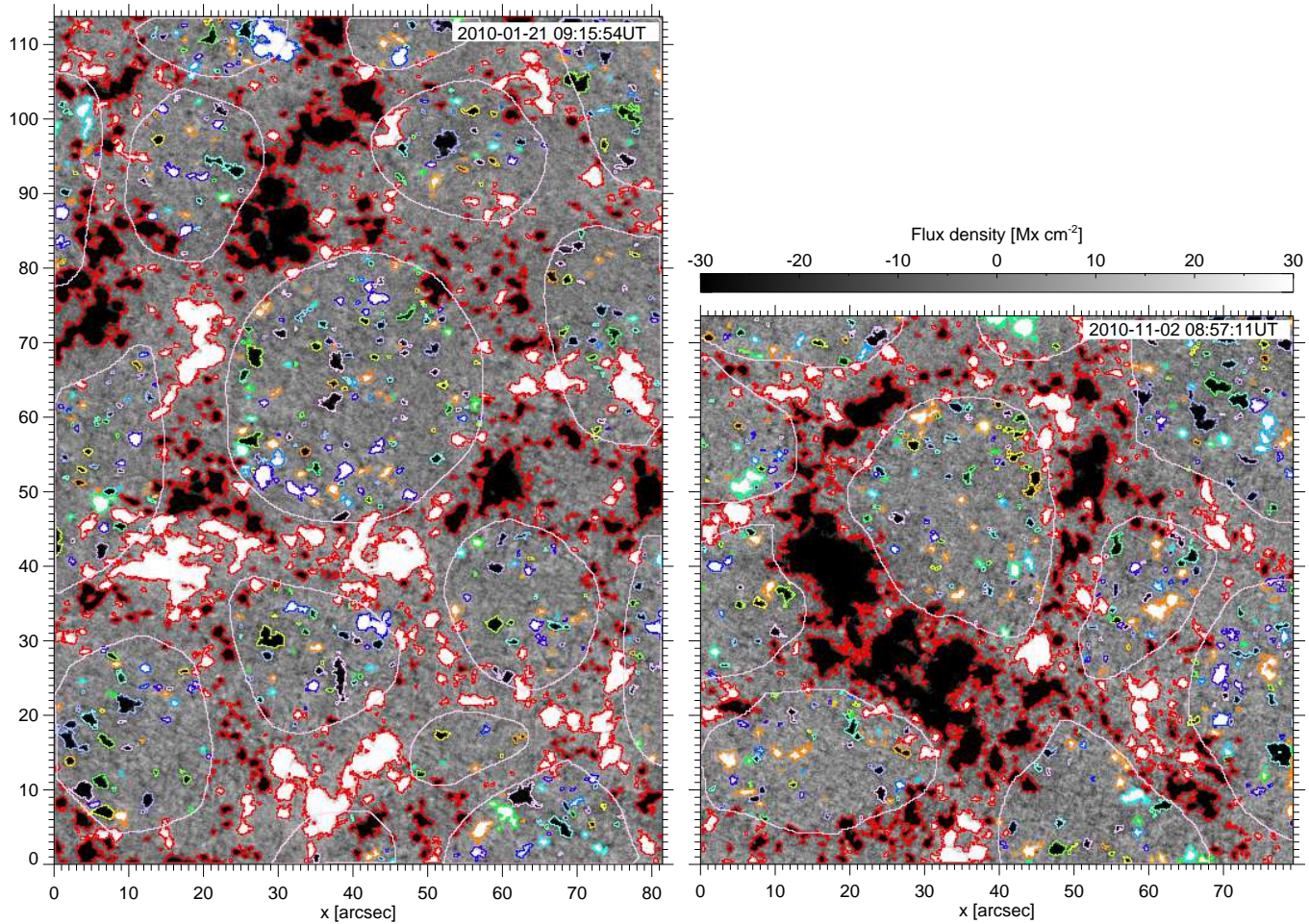}}
\end{center}
\vspace*{1em}
\caption{Individual magnetograms from data sets 1 ({\em left}) and 2
  ({\em right}). Both images are saturated at $\pm30$~Mx~cm$^{-2}$.
  Pink contours outline the boundaries of supergranular cells as
  deduced from an LCT analysis. They separate NE and IN regions. Red
  contours show the NE elements identified by our method. IN elements
  are displayed in colors other than red.}
\label{fig1}
\end{figure*}

The individual filtergrams were aligned and trimmed to their common
FOV. This step reduced the area of the observed regions to
$82\arcsec\times113\arcsec$ and $80\arcsec\times74$\arcsec,
respectively. Solar rotation was compensated during the data
acquisition, so as to monitor the evolution of the same quiet Sun
region for as long as possible. Both sequences contain several
supergranular cells and NE areas (see Figure~\ref{fig1}).

The observations were corrected for dark current and flatfield
following standard procedures. To generate gain tables we averaged all
the intensity filtergrams of each sequence. From the corrected
circular polarization and intensity images we calculated longitudinal
magnetograms as $M=0.5\times(V_{\rm b}/I_{\rm b}-V_{\rm r}/I_{\rm
  r})$, where r and b denote the red and blue line wings. 
The magnetogram signal was converted into magnetic flux density as
$\phi=C M$. We determined the calibration constant to be $C= 9160$~Mx
using the weak field approximation and the numerical value of
$1/I(\lambda) \, {\rm d} I(\lambda)/{\rm d}\lambda$ derived from the
intensity profile of the \ion{Na}{1} 589.6~nm line in the FTS atlas
\citep{fts_atlas}.
In addition, we computed Dopplergrams as
$D= (I_{\rm b}-I_{\rm r})/(I_{\rm b}+I_{\rm r})$. The Dopplergrams
were transformed into line-of-sight velocities applying a linear
relationship obtained by shifting the atlas profile by known amounts.

To remove systematic changes of the magnetogram signal resulting from
the Hinode orbital variations, we calculated the distribution of
magnetic flux densities for each frame of the sequence. The core of
the distribution was fitted with a Gaussian assuming that it
represents photon noise. Any deviation of the central position of the
Gaussian from zero was subtracted from all the pixels of the
corresponding frame. The observed offsets are periodic and show rms
values of less than 1.4~G.

Five minute oscillations were removed from the magnetogram and
Dopplergram time sequences by applying a subsonic filter
\citep{1989ApJ...336..475T, 1992A&A...256..652S}. The noise of the
magnetograms was $\sigma=6$ Mx cm$^{-2}$ at that stage, as computed in
a region without solar signals. To reduce the noise even further, we
spatially smoothed the magnetograms using a 3$\times$3 kernel of the 
form
\begin{equation}
k = \frac{1}{16} \left( \begin{array}{ccc} 
    1 & 2 & 1 \\
    2 & 4 & 2 \\
    1 & 2 & 1 
\end{array}
\right).
\end{equation}
This corresponds to a truncated Gaussian distribution with a full
width at half maximum of 2 pixels, which is smaller than the size of
the magnetic features studied in the present paper. The result is a
minimal degradation of the spatial resolution but a significant
improvement of the noise down to $\sigma=4$~Mx~cm$^{-2}$. Such a low
noise level permits a more reliable identification of the weakest IN
magnetic features and their association between consecutive
magnetograms.

\section{Method}
\label{method}
IN magnetic elements transfer their flux to the NE through
interactions with NE features. To study how they contribute to the
maintenance of the NE, we have to identify and track magnetic patches
in the magnetogram time sequences, classify them as IN or NE elements,
look for the IN features that interact with NE patches, and determine
how much flux they add to or remove from the NE. These steps are described
in detail below.

\subsection{Identification and tracking of magnetic features}
We use the YAFTA code\footnote{YAFTA (Yet Another Feature Tracking
  Algorithm) is an automatic tracking code written in IDL and can be
  downloaded from the author's website at
  http://solarmuri.ssl.berkeley.edu/~welsch/public/software/YAFTA.}
\citep{DeForest} to detect and track magnetic patches in the
magnetogram time sequences. The detection algorithm is based on the
clumping method, which groups together all contiguous like-polarity
pixels with absolute flux densities above a specified threshold and
marks them as a unique element \citep{Parnell2009}. We set a flux
density threshold of 12~Mx cm$^{-2}$ (3$\sigma$), a minimum size of 4
pixels, and a duration of at least 2 frames to consider one such clump
as a real magnetic feature.

Magnetic elements are in constant motion and often interact with other
features, gaining or losing flux, until they eventually disappear from
the solar surface. The following processes need to be considered to
track their evolution:
\begin{enumerate}
\item {\em In-situ appearance}, the process by which new flux elements
  become visible in the magnetograms. All the elements are
    counted individually, regardeless of their nature. Thus, the
    footpoints of emerging bipoles are taken to be independent
    entities in the same way as other unipolar patches.
\item {\em Merging}, or the coalescence of two or more elements of the
  same polarity into a larger structure. In this process, only the
  largest element survives, in the sense that it is the one tracked in
  the following frame. The smaller elements are considered disappeared
  features. The flux they contain is incorporated into the surviving
  element.
\item {\em Fragmentation}, the opposite of merging, occurs when an
  element splits into two or more smaller features (children).
\item {\em Cancellation}, or the disappearance of a magnetic element
  in the vicinity of an opposite-polarity feature. Through
  cancellations, magnetic flux is removed from the photosphere. 
\item {\em In-situ disappearance}, the process whereby magnetic
  elements disappear from the solar surface without an obvious
  interaction with any other feature.
\end{enumerate}

YAFTA assigns a unique label to each feature. These labels allow us to
trace back the history and origin of the elements detected in the
magnetograms. However, a careful analysis is needed because individual
elements may change their labels several times due to interactions.
For example, features lose their labels when they merge with a
stronger flux patch, cancel completely with an opposite-polarity
element, or disappear in situ. Likewise, the fragments of a feature
get new labels except the stronger one, which maintains the label of
the parent. Our strategy is to keep track of all the interactions to
avoid counting the same element more than once. We also correct the
YAFTA results in special cases involving complex interactions. Such
cases are described in Section~\ref{contribution}.

\subsection{Separation of NE and IN regions}
To identify NE and IN regions, some authors use criteria based on the
flux density and/or size of the magnetic elements. However, the
separation is difficult because the QS flux distribution is continuous
and does not show a clear boundary between the IN and NE
\citep{Hagenaar2003, ThorntonParnell, Zhou2013}. To avoid the
uncertainties associated with the use of a fixed flux threshold, we
define the IN as the interior of supergranular cells. The space beyond
the IN is the NE.

Supergranular cells are found using the Dopplergram time sequences.
We divide them in 2~h intervals and apply Local Correlation Tracking
\citep[LCT;][]{November} to each subsequence. The LCT algorithm
calculates horizontal velocities by comparing small subfields in two
successive images. The subfields are defined by a Gaussian tracking
window of full width at half-maximum of 60 pixels ($9\farcs6$). With
this criterion we suppress small-scale convective patterns, keeping
only the large-scale supergranular flows. From the resulting horizontal 
velocity field we calculate divergence maps. IN regions are identified by 
tracking the evolution of passively advected tracers (corks) on the divergence 
maps. Assuming that the NE flux is located at the edges of supergranules, we 
manually define the IN following the cork distributions and avoiding strong 
magnetic flux structures. To obtain a smooth transition of the cell boundaries 
from one subsequence to the next we use linear interpolation.

Figure~\ref{fig1} shows examples of separated NE and IN regions for
data sets 1 and 2, with the pink contours representing the
boundaries between the IN and the NE.

\begin{figure*}[t]
\begin{center}
\resizebox{1\hsize}{!}{\includegraphics[]{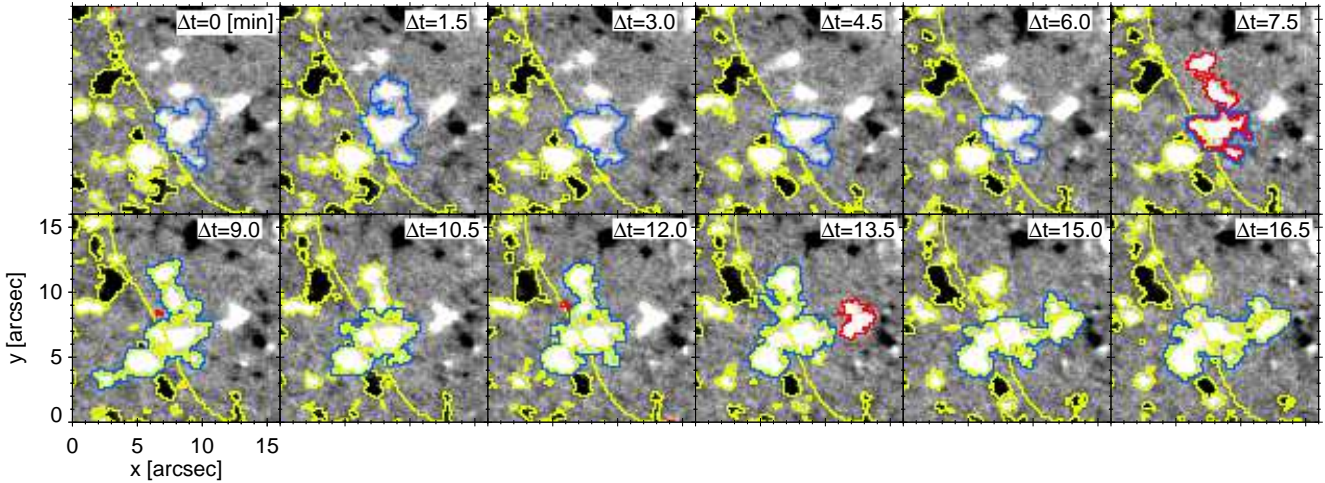}}
\end{center}
\vspace*{-1em}
\caption{Example of merging events. An IN element detected by YAFTA as
  a single feature (blue contours) merges with a NE patch at $\Delta
  t=9.0$ min and becomes a NE feature itself. NE elements are
  indicated with yellow contours. This specific IN patch undergoes
  several mergings with other IN patches during the sequence,
  revealing the full complexity of interactions between magnetic flux
  concentrations. For example, two smaller IN features to the North
  merge with it at $\Delta t = 9.0$~min.  The same happens with
  another IN element to the West at $\Delta t = 15.0$ min. The red
  contours mark the boundaries of the IN patch right before merging
  with the NE, as well as the boundaries of the other IN elements that
  will become part of it. The flux they enclose is taken to be the
  contribution to the NE of all the IN patches participating in the
  process. Note that the small IN elements that merge with the main IN
  feature at 9 min are assimilated into it. Therefore, we mark them as
  NE patches from that moment on and do not change their character
  when they separate from their parent at 10.5 and 15.0 min. }
\label{merging}
\end{figure*}

\subsection{Identification of IN and NE elements}

Magnetic features that appear in the interior of supergranular cells
(i.e., inside the boundaries defined above) are taken to be IN
elements. Elements born by fragmentation of an IN element are also
classified as IN patches. IN elements keep their character unless they
merge with NE features, in which case they become part of the NE and
are tagged accordingly.

The identification of NE patches is more involved. In the first frame,
elements that have their flux-weighted centers (FWC) inside of a NE
region are considered to be NE features. Also, only in the first
frame, elements that are touching a NE region are counted as NE
elements because we do not know their history.

For all $i>1$ frames, magnetic features are marked as NE elements if
they overlap more than 70$\%$ with NE structures from the previous
$i-1$ frame or if their FWCs are located inside any of the NE
patches detected in the $i-1$ frame.

When an element appears by fragmentation, it is considered to belong
to the NE if its parent is already marked as a NE structure. At any
time, an element that appears in situ inside the NE is considered 
to be part of the NE.

\subsection{Calculation of the IN contribution to the NE flux}
\label{contribution}
Merging and cancellation are the only two processes through which 
IN elements can modify the flux of the NE. These processes involve 
a direct interaction between IN and NE elements. IN features that 
merge with NE patches add flux to the NE, while those canceling with
opposite-polarity elements remove flux from the NE. Their 
respective contributions are determined as follows. 

\begin{figure*}[t]
\begin{center}
\resizebox{1\hsize}{!}{\includegraphics[]{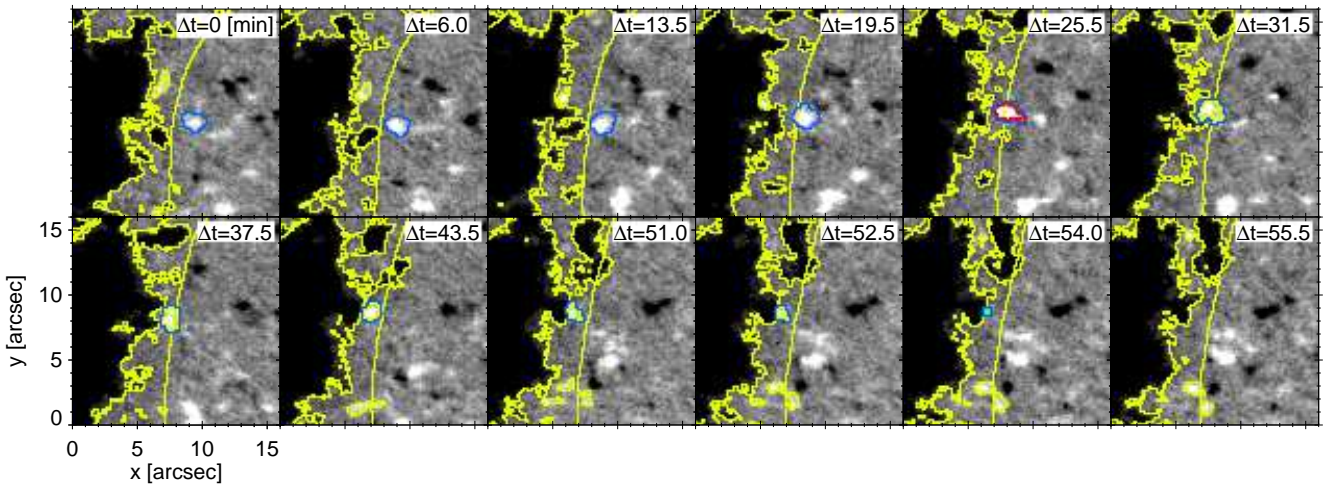}}
\end{center}
\vspace*{-1em}
\caption{Example of an IN element (blue contours) canceling with NE
  patches (yellow contours). The contours are those provided by YAFTA,
  except that the blue ones have been expanded by 1 pixel for
  visualization purposes. The red contours mark the flux used to
  compute the contribution of the IN element to the NE. The green
  contour at $\Delta t=54$~min outlines the canceling IN element in
  the last frame where it is visible.}
\label{cancellation}
\end{figure*}

\subsubsection{Contribution through merging events} 

When an IN element merges with a NE patch, the flux it contributes to
the NE is calculated in the frame where it is visible as an individual
feature for the last time. We consider only the pixels of the IN
element that do not overlap with any of the NE patches of the previous
frame, to avoid including NE flux due to fragments that YAFTA may have
erroneously added to the IN feature.

A detection problem arises when IN flux from the previous frame
converts into NE flux in the current frame, without ever being
recognized by YAFTA as a unique magnetic element. This happens when
some portion of an IN element fragments from its parent and merges
with a NE patch between two consecutive frames. To find and count such
events, we calculate the difference between logical masks of NE
elements in frame $i$ and frame $i-1$. All pixels different from zero
in the difference mask are assigned the flux densities they had in
frame $i$. On such ``partial'' magnetogram we apply the downhill
method. This method finds all contiguous like-polarity pixels above a
given threshold that lie on the same flux peak \citep{WelschLongcope}.
Every flux structure larger than 16 pixels exceeding a signal
threshold of 3$\sigma$ is taken to be a magnetic element. Finally, we
dilate these features to the nearest neighboring pixel. If they
overlap the NE mask from the current frame and their FWCs are not
located inside of the NE region, they are counted as contributing IN
fragments that were not detected by the code.

A similar situation occurs when an IN element merges with a NE
fragment which is not detected by YAFTA as an individual feature
because the fragmentation and the merging take place in the same
frame. For YAFTA, the IN element continues to be an IN element of
increased flux and size. If the NE fragment had more flux than the IN
patch, however, their common FWC will more likely be located inside
the NE feature of the previous frame from which the fragment detached,
and our method will recognize the entire element as a NE patch. To
evaluate the IN contribution in this case, and reject the NE portion
of the flux, we apply the downhill method described above.

A complex example of an IN element merging with NE patches is shown in
Figure~\ref{merging}. This element interacts with other IN features
during the sequence (e.g., at $\Delta t = 9.0$ and 15.0~min). The red
contours indicate the flux contributed to the NE by each of the IN
elements.

\subsubsection{Contribution through cancellation events}

Cancellation events are found by searching for all closely located,
opposite-polarity elements that disappear in the current frame. More
precisely, we take an IN element at the moment of disappearance,
dilate its border by 2 pixels and, if it overlaps a NE patch of
opposite polarity, count the IN element as a canceling flux feature.
To determine how much flux it removes from the NE, we go back in time
and take the total flux of the IN element at the moment when the
cancellation started (i.e, when the IN patch touched the
opposite-polarity NE element for the first time). This flux is
corrected for all the changes caused by the merging and fragmentation
processes that might happen during the cancellation event.

The same procedure is repeated for NE elements that disappear by
cancellation with IN elements. The latter may survive the process, 
but they remove flux from the NE. Their contribution is taken to be 
the flux of the NE patch at the beginning of the cancellation.

In Fig.~\ref{cancellation} we show an example of an IN feature
canceling with NE patches and how the interaction is interpreted.

\subsubsection{Practical implementation and limitations}

The identification of IN elements merging or canceling with NE
features is carried out in the following way. In frame $i$ we examine
all the NE and IN elements that lose their labels and disappear as
individual entities. We check whether they merged or canceled with
another element. Only if the interaction involves an IN element and a
NE patch do we count the former as contributing to the NE. In the case
of mergings, the flux carried by the IN patch in frame $i-1$ is
assumed to be transferred to the NE in frame $i$. In the case of
cancellations, the initial flux of the element that disappears
completely is taken to be the contribution of the IN to the NE in
frame $i$.

This method covers the situation in which IN elements merge with weak
NE fragments. Being dominant in flux they keep their labels, but they
must be considered part of the NE from that moment on (as well as
their children in case of fragmentation). Therefore, we tag them as
NE patches, except when they are inside the IN. This is done for
computational convenience---otherwise they would quickly turn all IN
patches into NE features through mergings with other elements inside
the supergranular cell. IN features not tagged as NE patches are
considered to contribute to the NE if they enter a NE region or end
their lives interacting with a NE patch.

The procedure just described determines the flux transferred to the NE
taking into account only the first interaction. IN elements that
cancel completely remove flux from the NE and cannot undergo further
interactions, so their contribution is estimated correctly. However,
in the case of mergings our calculations provide only an upper limit.
The reason is that an IN element may first merge with a NE patch and
later cancel with another NE patch, so that part of it would actually
remove rather than add flux to the NE. Unfortunately, once the
element loses its identity because of the merging there is no way to
know whether this occurs at all.

\begin{deluxetable}{lrr}
  \tablecolumns{3} 
 \tablewidth{\columnwidth} 

\tablecaption{Magnetic elements
    detected by YAFTA} \tablehead{ \colhead{ \hspace*{10pc}} & \colhead{Data set 1} &
    \colhead{Data set 2}} \startdata
  Total number of elements & 265\,933 & 316\,097 \\

  IN regions & & \\
  \hspace{1em} Total IN elements & 160\,859 & 215\,775\\
  \hspace{1em} Positive IN elements  & 80\,293 & 108\,902 \\
  \hspace{1em} Negative IN elements & 80\,566 & 106\,873 \\
  \hspace{1em} Mean effective size & $1\farcs0$ & $1\farcs0$ \\
  \hspace{1em} Mean absolute flux  $[\times10^{16}\mbox{~Mx}]$ &  12.7 & 13.6 \\
  \hspace{1em} Mean flux $[\times10^{16}\mbox{~Mx}]$ & $-1.1$ & $-0.08$ \\

  NE regions & & \\
  \hspace{1em} Total NE elements & 105\,074 & 100\,322\\
  \hspace{1em} Positive NE elements & 50\,777 & 43\,065\\
  \hspace{1em} Negative NE elements & 54\,297 & 57\,257 \\
  \hspace{1em} Mean effective size & $1\farcs8$ & $2\farcs0$ \\
  \hspace{1em} Mean absolute flux  $[\times10^{16}\mbox{~Mx}]$    & 71.7 & 123.3 
\\
  \hspace{1em} Mean flux $[\times10^{16}\mbox{~Mx}]$ & $-7.6$ & $-78.6$ 

\enddata
\label{table2}
\end{deluxetable}

\section{Results}
\label{results}

\begin{figure*}[t]
\begin{center}
\resizebox{1\hsize}{!}{\includegraphics[]{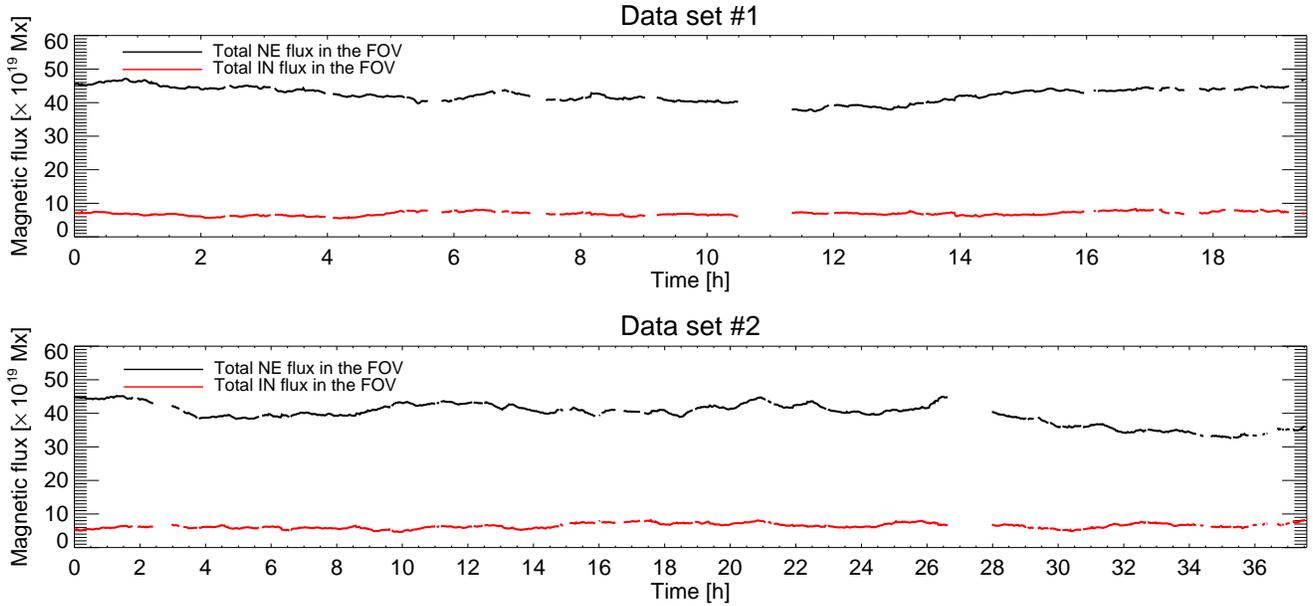}}
\end{center}
\vspace*{-2em}
\caption{Temporal evolution of the total flux of NE and IN regions
  (black and red curves) in data sets 1 and 2.}
\label{total_NEandIN_flux}
\end{figure*}

Using YAFTA, we followed the evolution of the magnetic flux patches
visible in the entire FOV. In total we detected 265\,933 elements in
data set 1 and 316\,097 in data set 2. Table~\ref{table2} provides
detailed information broken down into NE and IN regions. IN patches
are noticeably more abundant than NE patches, accounting for more than
60\% of all the elements. Such a result is expected from their larger
appearance rates and shorter lifetimes. In the data sets used here, IN
elements have a mean absolute flux of about $13 \times 10^{16}$~Mx.
Their mean flux is less than approximately $10^{16}$~Mx, indicating
that the two polarities are nearly balanced. By contrast, NE elements
are much stronger, with mean absolute fluxes of $72 \times 10^{16}$~Mx
in data set 1 and $123 \times 10^{16}$~Mx in data set 2. Both time
sequences exhibit a pronounced NE flux imbalance that may reach up 
to $-79 \times 10^{16}$~Mx.

The figures in Table~\ref{table2} do not reflect the number of unique
elements in the magnetograms because many of them are counted multiple
times by YAFTA. Our analysis is not affected by this behavior because
we are interested only in IN features interacting with NE elements. If
they survive the interaction, we label them as NE elements to avoid
their contributing more than once to the NE flux.

\subsection{Temporal evolution of IN and NE fluxes}

Figure~\ref{total_NEandIN_flux} shows the temporal evolution of the
total NE and IN fluxes for the two data sets analyzed here. These
values have been calculated adding the absolute fluxes of all the
detected NE and IN magnetic elements, respectively. Gaps in the curves
are the result of transmission problems and data recorder overflows.
Other short interruptions coincide with magnetograms that contain
recovered stripes\footnote{Lost telemetry packets produce empty stripes in the 
data. Whenever possible, we recovered them using only one line wing to compute 
the missing magnetogram signals, or the closest magnetogram in the sequence if 
the two wings were affected by telemetry problems. As a result, these areas 
have larger noise levels and are not used in the figures.}.

Both the NE and IN fluxes remain stable over time, showing only small
fluctuations (less than 8\% and 12\% rms, respectively). This
indicates that the NE and IN have reached a steady state. Individual
supergranular cells are appearing, evolving, and disappearing in the
time sequences, but the FOV is sufficiently large as for the results
not to depend on the variations of single cells.

The total NE and IN fluxes in data set 1 are on average $42.5 \times
10^{19}$~Mx and $6.9 \times 10^{19}$~Mx, respectively. For data set 2,
the corresponding values are $39.8 \times 10^{19}$~Mx and $6.5 \times
10^{19}$~Mx. Thus, 14\% of the total QS (NE$+$IN) flux is in the form
of IN elements. The remainder is provided by the NE, with a total flux of 
$5.5-8.0 \times 10^{23}$~Mx over the entire Sun. The minimum and maximum 
contributions of the IN to the QS flux observed in our data sets are 10\% and 
20\%, respectively.

\begin{figure*}[t]
\begin{center}
\resizebox{1\hsize}{!}{\includegraphics[]{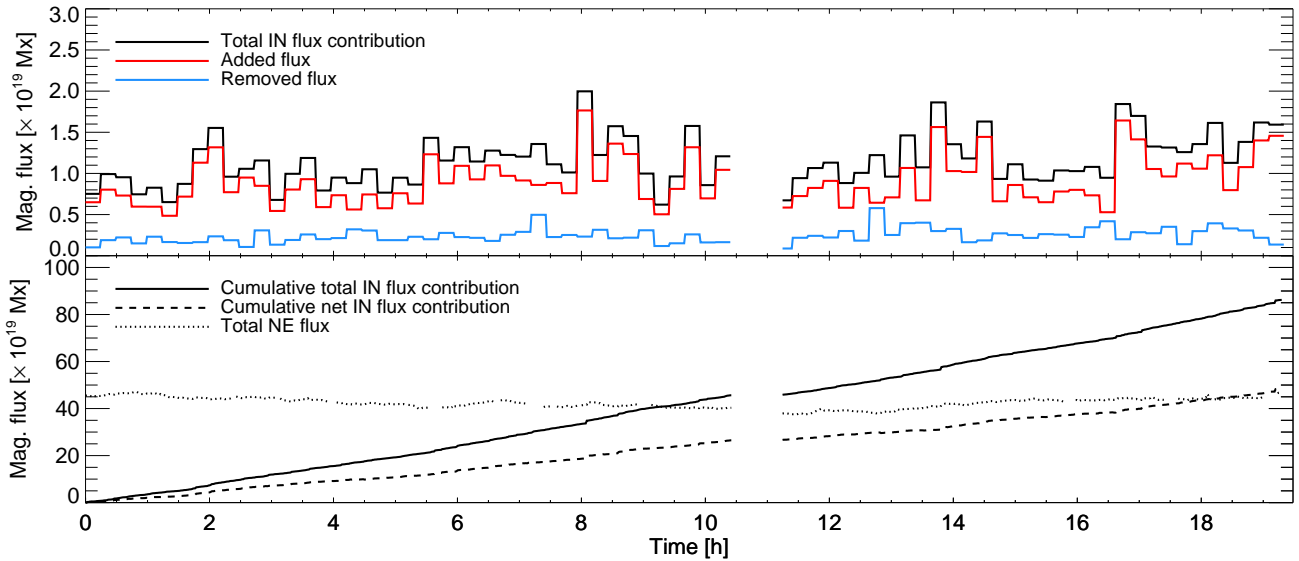}}
\end{center}
\vspace*{-2em}
\caption{Contribution of the IN to the NE in the full FOV of data
  set~1.  We use 15 minute bins to represent the data. {\em Top
    panel:} total contribution (black), defined as the sum of the
  absolute flux that the IN adds to the NE through mergings (red) and
  the absolute flux that the IN removes from the NE through
  cancellations (blue). The average total contribution per 15 min bin
  is $1.15 \times 10^{19}$~Mx, the average added flux is $0.91 \times
  10^{19}$~Mx, and the average removed flux is $0.24 \times
  10^{19}$~Mx. {\em Bottom panel:} Cumulative contribution of the IN
  to the NE flux. The solid line represents the total IN flux
  transferred to the NE (added plus removed absolute flux), while the
  dashed line gives the net contribution of the IN to the NE (added
  minus removed flux). The total NE flux in the entire FOV is
  indicated by the dotted line.}
\label{fig5}
\end{figure*}

\begin{figure*}[t]
\begin{center}
\resizebox{1\hsize}{!}{\includegraphics[]{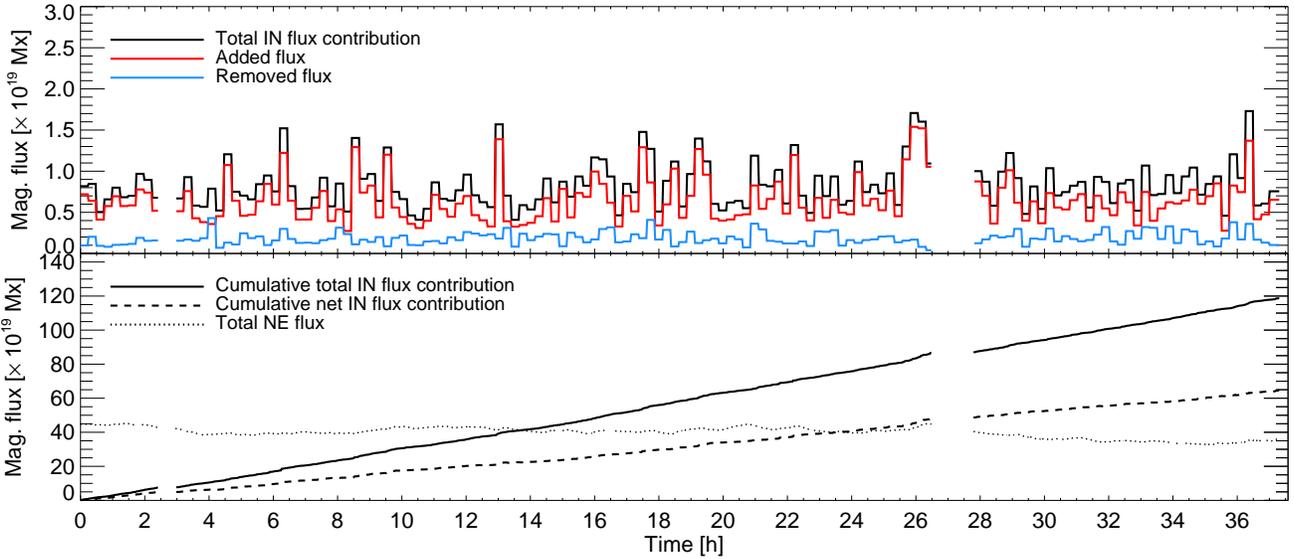}}
\end{center}
\vspace*{-2em}
\caption{Same as Figure \ref{fig5}, but for data set~2. The average
  total IN contribution per 15 min bin is $0.82 \times 10^{19}$~Mx,
  the average added flux is $0.64 \times 10^{19}$~Mx, and the average
  removed flux is $0.18 \times 10^{19}$~Mx.}
\label{fig6}
\end{figure*}

\begin{figure*}[t]
\begin{center}
\resizebox{1\hsize}{!}{\includegraphics[]{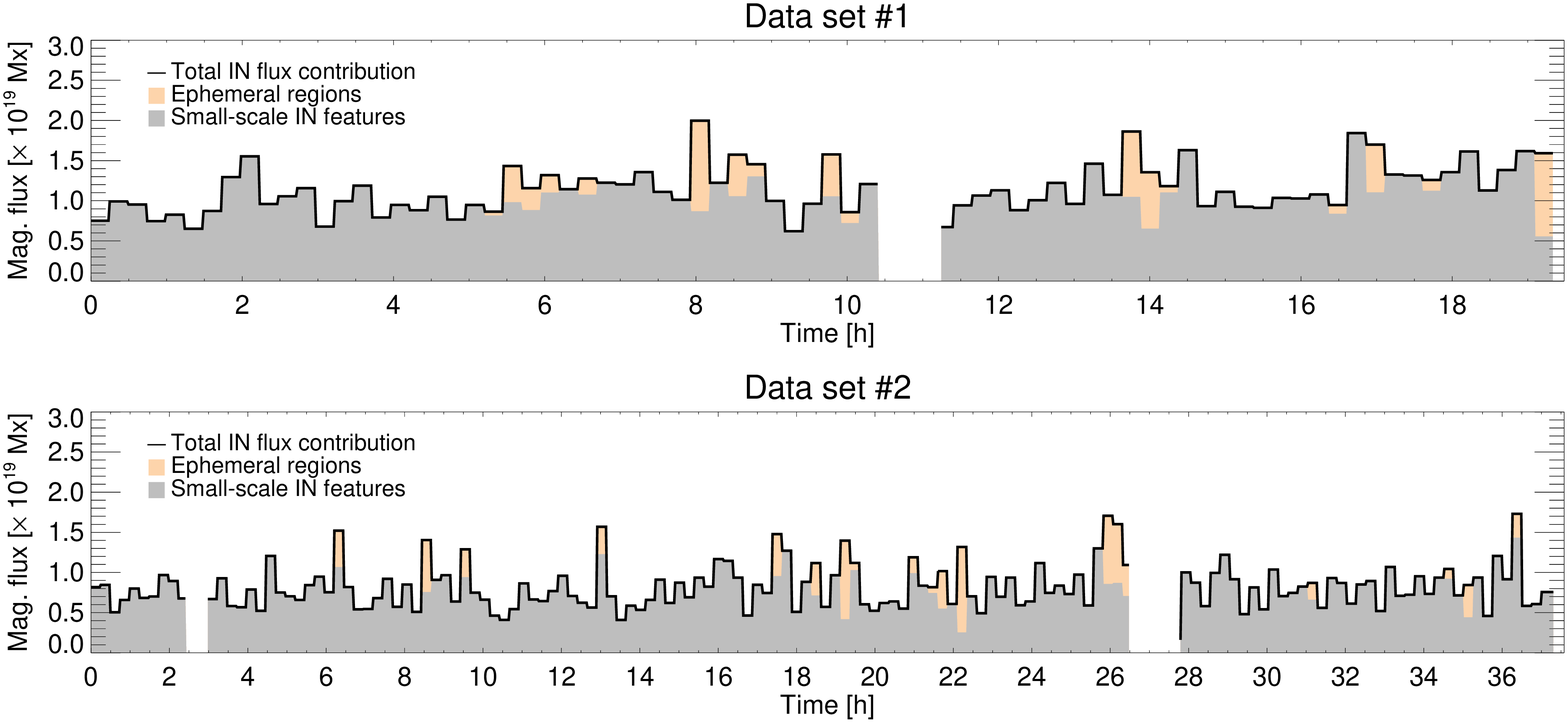}}
\end{center}
\vspace*{-1em}
\caption{IN contribution to the NE flux in the form of isolated IN
  elements (gray shaded areas) and ER patches (orange shaded areas).
  The total IN contribution is represented by the black curves. }
\label{fig7}
\end{figure*}

\subsection{Contribution of the IN to the NE flux}

The top panel of Figure \ref{fig5} shows how much flux the IN
transfers to the NE as a function of time in data set~1. The two
possible contributions are separated, namely flux added to the NE
through mergings (red curve) and flux removed from the NE through
cancellations (blue curve). Their sum is indicated by the black curve.
The data have been binned over 15 minutes. Each bin represents the IN
flux contributed to the NE during that period of time.

As can be seen, the total IN contribution to the NE shows some
variations over the 19.5~h of observation, but it is relatively
constant at about $4.6 \times 10^{19}$~Mx~h$^{-1}$ in the full FOV, or
$1.4 \times 10^{24}$~Mx~day$^{-1}$ over the entire solar surface.
Clearly, the IN is more efficient in injecting flux than in removing
it. Merging of like-polarity patches is the dominant process and
involves 3.8 times more flux than cancellations.

In the bottom panel of Figure \ref{fig5} we show the cumulative IN
contribution to the NE considering all the flux deposited in the NE
(solid line) and the net flux gained by the NE (dashed line). The
former is the total IN flux that actually interacts with NE patches,
independently of the final outcome of the interaction. The latter is
the net flux added to the NE (mergings minus cancellations).

The slope of the two curves does not change with time, indicating that
the IN supplies flux to the NE at a nearly constant rate. Very
interestingly, in only $\sim$9.5 hours the NE receives from the IN as
much flux as it contains. Most---but not all---of that flux is
incorporated into the NE. Considering only the net flux that remains
in the NE (dashed line), the IN needs some 18~h to supply as much flux
as is present in the NE at any one time.

The IN contribution to the NE flux in data set~2 is shown in
Figure~\ref{fig6}. Just like in the first data set, mergings dominate
over cancellations by a factor of 3.6, the result being that the IN
adds flux to the NE continually (top panel). On average, the IN
contributes a total of $3.3 \times 10^{19}$~Mx~h$^{-1}$ to the NE in
the observed FOV, or $1.6 \times 10^{24}$~Mx~day$^{-1}$ over the entire
solar surface. The cumulative contributions are shown in the bottom
panel, both for the total flux transferred to the NE (solid line) and
for the net flux which is actually incorporated into the NE (dashed
line). The dotted line represents the total NE flux in the FOV.
According to these curves, the IN transfers as much flux as the NE
contains in 13.5~h, or in some 24~h considering only the net
contribution to the NE.

\begin{figure*}[t]
\begin{center}
\resizebox{1\hsize}{!}{\includegraphics[]{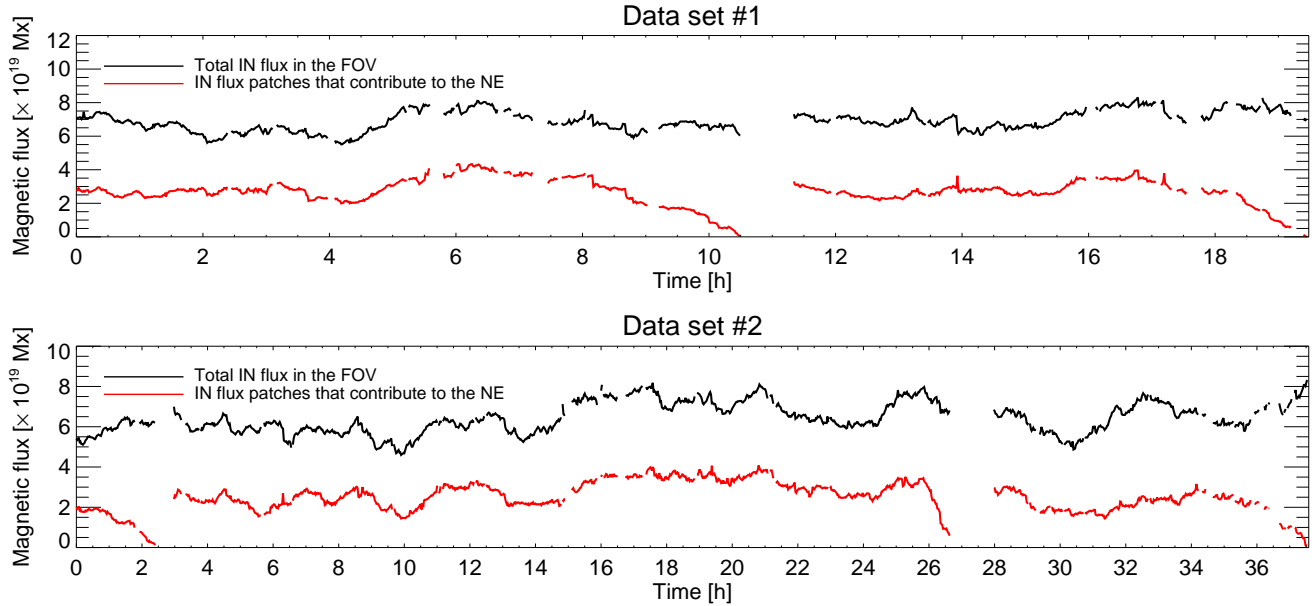}}
\end{center}
\vspace*{-1em}
\caption{Fraction of the total IN magnetic flux that converts into NE
  flux, for data sets 1 and 2 (top and bottom, respectively). The red
  curves show the instantaneous flux carried by the IN patches that
  will convert to the NE at any later time. For reference, the black
  curves represent the IN flux in the entire FOV.}
\label{frac}
\end{figure*}

Thus, the IN continually supplies flux to the NE. Since the total
NE flux does not change with time, flux must be removed from the NE at
a similar rate through cancellations and in-situ disappearances.
However, a detailed analysis of such processes is beyond the scope of
this paper, as it would require careful interpretation of the YAFTA
results in the extremely crowded and dynamic regions of the NE. First
attempts to quantify the interactions between magnetic elements in
the NE have been presented by \cite{2012ApJ...752..149I}.

\subsection{Contribution of ERs to the NE flux}

The current paradigm is that ERs provide the flux needed to maintain
the network. Our very long magnetogram sequences can be used to
determine the percentage of the IN flux transferred to the NE that is
due to ERs. Detecting ERs in the Hinode magnetograms is difficult because YAFTA 
does not recognize the presence of large-scale structures. To circumvent this 
problem, we manually identified ERs as clusters of mixed-polarity flux patches 
appearing together and separating from each other with time. To qualify as an 
ER, at least one element in the cluster had to reach a flux content of $3 
\times 10^{18}$~Mx, the threshold set by \cite{Hagenaar2003}. The individual ER 
patches were then tracked and counted as contributing to the NE if they had 
interactions with NE patches. We detected five ERs in data set 1 and nine in 
data set 2, roughly consistent with the appearance rate of 430 per hour 
estimated by \citet{Chae} over the entire solar surface.

The black curves in Figure~\ref{fig7} show the total IN contribution
to the NE already displayed in the top panels of Figures~5 and 6. For
each 15 minute bin we separate the flux transferred to the NE by
isolated IN patches (gray shaded areas) and ER patches (orange shaded
areas). 

As can be seen, ERs play only a minor role in the flux transfer from
the IN to the NE. They inject about $1.1 \times 10^{23}$~Mx~day$^{-1}$
into the IN, but this is less than $\sim$8\% of the total IN flux that
reaches the NE in the two data sets. Thus, despite the large amount of
flux they carry, ERs are not the main contributors to the NE flux.
More than 90\% of the flux comes from weak, isolated IN patches.

\subsection{Fraction of total IN flux transferred to the NE}
After appearing in the interior of supergranular cells, IN elements
migrate toward the cell boundaries where the NE resides. However, not
all IN elements end up in the NE, since many of them disappear in situ
or cancel with other IN elements before reaching the NE. Here we study
what fraction of the IN flux actually makes it to the NE.

The black curves in Figure~\ref{frac} show the temporal evolution of
the IN flux in the entire FOV of data sets 1 and 2. The total flux
carried by the IN patches that interact with the NE at any later time
is indicated with the red curves. For this calculation we take into
account all fragmentations and partial cancellations that the elements
may undergo before interacting with the NE. Close to the end of each
continuous observing periods there are artificial drops of the IN flux
contributing to the NE. They reflect the lack of information about the
evolution of IN elements during the data gaps. Some of the elements
may interact with the NE, but they are not counted because it is not
possible to identify them, producing drops in the curves.

On average, we find that 38\% of the IN flux generated in the interior
of supergranules eventually contribute to the NE. Both data sets show
very similar percentages. The rest of elements disappear inside the 
IN by cancellations or in-situ fading and never reach the NE (see Paper
II of this series by Go\v{s}i\'{c} \& Bellot Rubio, in
preparation). Thus, a significant fraction of the observed IN flux is
incorporated into the NE, emphasizing the importance of the IN for the
maintenance---and arguably also the existence---of the magnetic NE.

\section{Discussion and Conclusions}

We have analyzed magnetic flux elements in NE and IN regions using
long-duration, high-resolution Hinode/NFI magnetogram sequences. We
carefully processed the data to facilitate the tracking of magnetic
features and the identification of patches after interactions. These
measurements give us the opportunity to study for the first time the
evolution of QS magnetic fields in the range $10^{16}-10^{20}$ Mx on
scales from minute to days. After identifying NE and IN flux
concentrations, we followed them with an automatic tracking algorithm
to examine how the IN contributes flux to the NE. We calculated the
transfer of IN flux to the NE in a direct way, using the instantaneous
flux of the IN elements that interact with NE patches.

In our observations, the total IN and NE fluxes are very stable and
exhibit only small variations with time. IN regions are in nearly
perfect polarity balance. By contrast, NE regions show clear polarity
imbalances in both data sets. On average, 14\% of the QS flux is in
the form of IN elements. This value is consistent with the estimates
of \cite{Wang}. The instantaneous ratio, however, varies in the range
from 10\% up to 20\%. Our definition of NE and IN elements is based on
spatial location---derived from horizontal flow maps---rather than on
flux densities or sizes. This may have resulted in lower percentages
because some small-scale IN elements were probably labeled as NE
features. If flux density or size criteria had been used instead,
those elements would have more likely been recognized as IN elements.
However, as they appear close to NE patches in supergranular
downflows, it seems reasonable to mark them as NE fragments.

We have shown that IN elements continuously supply flux to the NE,
confirming the prediction by \citet{Lamb2008} that it should be possible
to observe interactions between NE and IN elements on spatial scales
not accessible to MDI.

We estimate that up to 38\% of the magnetic flux appearing inside
supergranular cells moves toward the edges of supergranules and
interact with the NE. IN features merge or cancel with preexisting NE
patches, adding or removing flux from the NE. The first process is
dominant, by a factor of 3.6--3.8. The IN contributes a total flux of
$\sim1.5 \times 10^{24}$~Mx~day$^{-1}$ to the NE. Thus, IN regions
can generate and transfer as much magnetic flux as is present in NE
regions in only 9--13~h. Taking into account that not all elements add
flux to the NE, the time needed by the IN to supply a net flux equal
to that present in the NE is 18-24 hours.

Our estimates of the IN contribution to the NE flux are more accurate
than those based on flux emergence or disappearance rates. This is
because we directly measure how much flux converts into NE flux or
cancels with NE elements instantaneously, whereas flux emergence and
cancellation rates include a fraction of elements that never interact
with the NE and therefore should not be counted. In our data sets,
the temporal scales on which the IN could supply as much flux as is
contained in the NE are shorter than the 36--72~h reported by
\citet{Schrijver}, but also significantly longer than the lower limit
of 1~h found by \cite{Hagenaar2008}. Both analyses used SOHO/MDI data,
and therefore could not see most of the IN flux.

Our observations made it possible to determine in a direct way the
actual contribution of ERs to the NE flux. To estimate the flux they
transfer to the NE we have considered as ERs all the clusters of
mixed-polarity patches emerging in the same area with at least one of
the patches reaching $3 \times 10^{18}$~Mx. During the approximately
58 hours of monitoring of the quiet Sun performed with the Hinode NFI
we see 14 ERs. We find that ERs transfer flux to the network at a rate
of $1.1 \times 10^{23}$~Mx~day$^{-1}$, which is only slightly larger
than the ER emergence rate of $7.2 \times 10^{22}$~Mx~day$^{-1}$
reported by \cite{Schrijver} but less than 8\% of the total IN
contribution to the NE found here. This entails a change of paradigm,
as most of the IN flux transferred to the NE seems to come from weak,
isolated IN elements and not from ERs.

The main result of this work is that NE regions receive a substantial
and constant flux inflow from the solar IN. Actually, small-scale IN
elements appear to be the main and most permanent source of flux for
the NE. Their origin however remains poorly understood. To shed light
on this issue, we will determine the flux appearance and disappearance
rates in the IN in Paper II of this series, while Paper III will be
devoted to study the modes of appearance of IN elements on the solar
surface.

\acknowledgments 

This paper is based on data acquired in the framework of the Hinode
Operation Plan 151, {\em ``Flux replacement in the solar network and
internetwork''}. We thank the Hinode Chief Observers for the efforts
they made to accommodate our demanding observations. Hinode is a
Japanese mission developed and launched by ISAS/JAXA, with NAOJ as a
domestic partner and NASA and STFC (UK) as international partners. It
is operated by these agencies in co-operation with ESA and NSC
(Norway). M.G. acknowledges a JAE-Pre fellowship granted by Agencia
Estatal Consejo Superior de Investigaciones Cient\'{\i}ficas (CSIC)
towards the completion of a PhD degree. This work has been funded by
the Spanish Ministerio de Econom\'{\i}a y Competitividad through
project AYA2012-39636-C06-05, including a percentage from European
FEDER funds. Use of NASA's Astrophysical Data System is gratefully
acknowledged.


\begin{thebibliography}{}


\bibitem[Brault \& Neckel(1987)]{fts_atlas} Brault, J. W., \& Neckel, H. 1987, Spectral Atlas of Solar Absolute Disk-averaged and Disk-Center Intensity from 3290 to 12510~\AA\/, ftp://ftp.hs.uni-hamburg.de/pub/outgoing/FTS-Atlas

\bibitem[Bellot Rubio \& Orozco Su\'{a}rez(2014)]{BellotOrozco} Bellot Rubio, L.R., \& Orozco Su\'arez, D. 2014, Living Reviews in Solar Physics, submitted

\bibitem[Centeno et al.(2007)]{Centeno} Centeno, R., Socas-Navarro, H., \& Lites, B.\ et al.\ 2007, \apj, 666, L137 

\bibitem[Chae et al.(2001)]{Chae} Chae, J., Martin, S.~F., \& Yun, H.~S.\ et al.\ 2001, \apj, 548, 497 

\bibitem[DeForest et al.(2007)]{DeForest} DeForest, C.~E., Hagenaar, H.~J., \& Lamb, D.~A.\ et al.\ 2007, \apj, 666, 576 

\bibitem[De Wijn et al.(2009)]{DeWijn} De Wijn, A.~G., Stenflo, J.~O., \& Solanki, S.~K.\ et al.\ 2009, \ssr, 144, 275 

\bibitem[Go{\v{s}}i{\'c}(2012)]{milan} Go\v{s}i\'c, M.\ 2012, Master Thesis, University of Granada (Spain)

\bibitem[Hagenaar (2001)]{Hagenaar2001} Hagenaar, H.~J.\ 2001, \apj, 555, 448 

\bibitem[Hagenaar et al.(2008)]{Hagenaar2008} Hagenaar, H.~J., DeRosa, M.~L., \& Schrijver, C.~J.\ 2008, \apj, 678, 541 

\bibitem[Hagenaar et al.(2003)]{Hagenaar2003} Hagenaar, H.~J., Schrijver, C.~J., \& Title, A.~M.\ 2003, \apj, 584, 1107

\bibitem[Harvey \& Martin (1973)]{HarveyMartin} Harvey, K.~L., \& Martin, S.~F.\ 1973, \solphys, 32, 389

\bibitem[Harvey et al.(1975)]{Harvey1975} Harvey, K.~L., Harvey, J.~W., \& Martin, S.~F.\ 1975, \solphys, 40, 87

\bibitem[Iida et al.(2012)]{2012ApJ...752..149I} Iida, Y., Hagenaar, H.~J., 
\& Yokoyama, T.\ 2012, \apj, 752, 149 

\bibitem[Kosugi et al.(2007)]{2007SoPh..243....3K} Kosugi, T., Matsuzaki, 
K., Sakao, T., et al.\ 2007, \solphys, 243, 3 

\bibitem[Lamb et al.(2008)]{Lamb2008} Lamb, D.~A., DeForest, C.~E., \& Hagenaar, H.,~J.\ et al. \ 2008, \apj, 674, 520

\bibitem[Lamb et al.(2010)]{Lamb2010} Lamb, D.~A., DeForest, C.~E., \& Hagenaar, H.,~J.\ et al. \ 2010, \apj, 720, 1405

\bibitem[Lites(2002)]{Lites2002} Lites, B.~W.\ 2002, \apj, 573, 431


\bibitem[Livingston \& Harvey(1975)]{LivingstonHarvey} Livingston, W.~C., \& Harvey, J.\ 1975, \baas, 7, 346

\bibitem[Martin(1990)]{Martin1990} Martin, S.~F.\ 1990, IAUS, 138, 129

\bibitem[Mart{\'{\i}}nez Gonz{\'a}lez \& Bellot Rubio(2009)]{MartinezLuis} Mart{\'{\i}}nez Gonz{\'a}lez, M.~J., \& Bellot Rubio, L.~R.\ 2009, \apj, 700, 1391

\bibitem[Meunier et al.(1998)]{Meunier} Meunier, N., Solanki, S.~K., \& Livingston, W.~C.\ 1998, \aap, 331, 771

\bibitem[November \& Simon(1988)]{November} November, N., \& Simon, W.~G.\ 1998, \apj, 333, 427

\bibitem[Orozco Su{\'a}rez et al.(2008)]{Orozco2008} Orozco Su{\'a}rez, D., Bellot Rubio, L.~R., \& del Toro Iniesta, J.~C.\ et al.\ 2008, \apjl, 481, L33 

\bibitem[Orozco Su{\'a}rez et al.(2012)]{Orozco2012} Orozco Su{\'a}rez, D., Katsukawa, Y., \& Bellot Rubio, L.~R.\ 2012, \apjl, 758, L38 

\bibitem[Parnell et al.(2009)]{Parnell2009} Parnell, C.~E., DeForest, C.~E., \& Hagenaar, H.~J.\ et al.\ 2009, \apj, 698, 75

\bibitem[S{\'a}nchez Almeida \&
Mart{\'{\i}}nez Gonz{\'a}lez(2011)]{2011ASPC..437..451S} S{\'a}nchez Almeida,
J., \& Mart{\'{\i}}nez Gonz{\'a}lez, M.\ 2011, ASP Conf.\ Series, 437, 451 

\bibitem[Scherrer et al.(1995)]{Scherrer} Scherrer, P.~H., Bogart, R.~S., \& Bush, R.~I.\ et al.\ 1995, \solphys, 162, 129 

\bibitem[Schrijver \& Harvey(1994)]{1994SoPh..150....1S} Schrijver, C.~J., \&
Harvey, K.~L.\ 1994, \solphys, 150, 1 

\bibitem[Schrijver et al.(1997)]{Schrijver} Schrijver, C.~J., Title, A.~M., \& Hagenaar, H.~J.\ et al.\ 1997, \apj, 487, 424 

\bibitem[Simon et al.(2001)]{Simon2001} Simon, G.~W., Title, 
A.~M., \& Weiss, N.~O.\ 2001, \apj, 561, 427 

\bibitem[Solanki(1993)]{Solanki} Solanki, S.~K.\ 1993, \ssr, 63, 1

\bibitem[Straus et 
al.(1992)]{1992A&A...256..652S} Straus, T., Deubner, F.-L., \& Fleck, B.\ 1992, \aap, 256, 652

\bibitem[Title(2000)]{Title} Title, A.~M.\ 2000, Philos. Trans. Roy. Soc. London A, 358, 657 

\bibitem[Title et al.(1989)]{1989ApJ...336..475T} Title, A.~M., Tarbell, 
T.~D., Topka, K.~P., et al.\ 1989, \apj, 336, 475 

\bibitem[Thornton \& Parnell(2010)]{ThorntonParnell} Thornton, L.~M., \& Parnell, C.~E.\ 2010, \solphys, 269, 13 

\bibitem[Tsuneta et al.(2008)]{Tsuneta} Tsuneta, S., Ichimoto, K., \& Katsukawa, Y.\ 2008, \solphys, 249, 167 

\bibitem[Wang \& Zirin(1987)]{WangZirin} Wang, H., \& Zirin, H.\ et al.\ 1987, \solphys, 115, 205

\bibitem[Wang et al.(1995)]{Wang} Wang, J., Wang, H., \& Tang, F.\ et al.\ 1995, \solphys, 160, 277

\bibitem[Wang et al.(2012)]{2012SoPh..278..299W} Wang, J., Zhou, G., Jin, C., \& Li, H.\ 2012, \solphys, 278, 299 

\bibitem[Welsch \& Longcope(2003)]{WelschLongcope} Welsch, B.~T., \& Longcope, D.~W.\ 2003, \apj, 588, 620

\bibitem[Zhou et al.(2010)]{Zhou2010} Zhou, G., Wang, J., \& Jin, C.\ 2010, \solphys, 267, 63z

\bibitem[Zhou et al.(2013)]{Zhou2013} Zhou, G., Wang, J., \& Jin, C.\ 2013, \solphys, 283, 273


\bibitem[Zirin(1985)]{Zirin1985} Zirin, H.\ 1985, AuJPh, 38, 961 

\bibitem[Zirin(1987)]{Zirin1987} Zirin, H.\ 1987, \solphys, 110, 101 


\end{thebibliography}
\end{document}